\documentstyle[prb,aps,epsf,twocolumn]{revtex}
\begin{document}

\draft
\twocolumn[\hsize\textwidth\columnwidth\hsize\csname@twocolumnfalse%
\endcsname

\title{Two-terminal conductance fluctuations in the integer quantum Hall regime}

\author{Chang-Ming Ho}

\address{Theoretical Physics, University of Oxford, 1 Keble Road, Oxford OX1 3NP, 
United Kingdom \cite{todai}}

\date{{\it To be published in} Physical Review B (Sep, 1999)}

\maketitle

\begin{abstract}
Motivated by recent experiments on the conductance fluctuations in mesoscopic 
integer quantum Hall systems, we consider a model in which the Coulomb interactions are 
incorporated into the picture of edge-state transport through a single saddle-point. 
The occupancies of {\it classical} localised states in the two-dimensional electron system 
change due to the interactions between electrons when the gate voltage on top of the device is 
varied. The electrostatic potential between the localised states and the saddle-point 
causes fluctuations of the saddle-point potential and thus fluctuations of 
the transmission probability of edge states. This simple model is studied numerically 
and compared with the observation.    
\end{abstract}
 

\vspace{0.05in}
]
\narrowtext

\section{Introduction}

Mesoscopic sample-to-sample fluctuations in the conductance, $G$, are universal in the 
sense that the deviation from its average, $(\delta G)^{2}=(G-\langle G \rangle)^{2}$, 
is always about the order of $(e^{2}/h)^{2}$ at very low temperature \cite{ucf-exp,washburn}. 
In zero or weak external magnetic fields, this behaviour can be understood in terms of the quantum phase 
coherence of diffusive electrons \cite{ucf}. This picture, however, is not expected to work in the 
quantum Hall regime since the trajectories of electrons are drastically deformed in strong magnetic 
fields. In this situation, the existence of extended states at sample edges makes the transport 
properties rather different from the weak-field case \cite{halperin}.  Indeed, magnetoconductance 
measurements on mesoscopic multiterminal devices \cite{simmons,maingeim,geim,SdH} seem to agree with 
this expectation. To be consistent with these observations, explicit inclusion of the 
edge-channel effects in theories seem to be necessary \cite{jainkivelson,kinaret,dima}.

The appearance of the bulk delocalised states when the Fermi level is between two quantised Hall 
plateaus makes a mesoscopic system in strong magnetic field even more interesting \cite{huckstein}. It 
is only recently, however, that experimental data on the statistical behaviour of the conductance in 
this regime has become available \cite{cobdenkogan}. Specifically, conductance fluctuations has been 
studied in mesoscopic $n$-channel Silicon Metal-Oxide-Semiconductor Field-Effect Transistors({\it Si}\ MOSFETs) 
\cite{andofowlerstern} with two terminals. In these systems, conductance is quantised (in units of $e^{2}/h$) 
at some values of the gate voltage, $V_{g}$, and the complication due to the Shubnikov-de Haas 
oscillation in multiterminal devices is absent in the two-terminal setting. It is found that the 
distribution of conductance below the first conductance plateau is almost uniform between $G=0$ and $e^{2}/h$. 
This is in clear contrast to the Gaussian distribution in the weak-field case \cite{cobdenkogan}. 
Theoretical studies have been carried out on models based on single-particle interference effect, 
{\it e.g.} tight-binding Hamiltonian \cite{ando}, 
Chalker-Coddington network model \cite{wangjovanovic,chofisher,cf-network}, and so on 
\cite{avishai} numerically and also a renormalisation group analysis \cite{galstyan}. 
Some capture this feature of the conductance distribution. 

More recently, Cobden and collaborators demonstrate further that there is a great difference 
between conductance fluctuations in the lowest few Landau levels 
and the weak field case \cite{cobden-prod,cobden-unpublished}. Using the same sets of {\it Si}\ MOSFET 
devices as described, they observed in strong fields strong correlations of fluctuation 
peaks and dips at different magnetic fields. 
This is achieved by studying explicitly the evolution of fluctuation peaks and dips with magnetic 
fields in different field regimes. Using a greyscale plot of $dG/dV_{g}$ as a function of $B$ and $V_{g}$ 
, each bright or dark line can be viewed as the `history' line of a peak or a dip of the 
conductance at different combinations of $V_{g}$ and $B$. The contrast between weak and 
strong magnetic fields is very clear (See Fig. 2 in Ref. 22). 

For weak magnetic fields, no pattern can be seen whatsoever. 
At $B \geq$10T, we can see peaks and dips of conductance 
fluctuations in a transition region between two successive conductance 
plateaus, which are wide and grey regions. The fluctuations `move' in gate voltage, 
in straight lines as the magnetic field is varied. This pattern means that there exist 
strong correlations between fluctuation peaks and dips at different gate voltages at high 
magnetic fields. A closer look reveals that there are two types of lines, 
with different slopes, in a transition region. The slope of one set of 
lines is parallel to that associated with the centre of the plateau below, 
while the slope of the other set of lines is parallel to that associated 
with the centre of the plateau above. 

The strong contrast between the fluctuations on the $V_{g}$-$B$ plots in weak and strong 
magnetic fields means that what occurs in two different field regimes is dramatically 
different. The straight lines appearing in the 
strong-field results makes it no longer appropriate to treat the phenomenon 
as a single-particle one. Thus those models 
\cite{ando,wangjovanovic,chofisher,cf-network,avishai,galstyan} 
we mentioned before are obviously not compatible with strong correlations of 
conductance fluctuations at different fields.    

In this paper, we consider a model taking into account the electron-electron interactions.
The conductance plateaus are understood in terms of the edge-state picture and the absence of 
backscattering between channels at opposite edges of 
the sample. In the transition region between plateaus, we assume that the paths 
followed by electron currents coming from two contacts (or terminals) percolate 
into the bulk as the chemical potential is increased and eventually become connected 
by tunneling through a single saddle point. In addition to the electrons in these edge states, 
we also take account of electrons in localised states. Conductance fluctuations arising from 
Coulomb interactions between both sets of electrons are treated in a simple way. 

As electrons are added to the system, the ones already in the localised states 
will rearrange due to the interactions. 
Therefore, the occupation of a given localised state can fluctuate several times 
between $0$ and $1$ when the chemical potential, $\mu$, is changed from $-\infty$ to $+\infty$. 
The total electrostatic potential between the localised states and the saddle point then also 
fluctuates as $\mu$ is varied. Because of the fluctuations of the saddle point potential, the tunneling
of percolation paths fluctuates as a function of the chemical potential. In our model, fluctuations 
of the conductance are purely due to these interaction effects. Numerical study of this model shows 
that the conductance between plateaus as a function of the chemical potential does indeed fluctuate 
significantly within the range between $G=0$ and $e^{2}/h$.

In the next section, we first develop our model, based on the transmission of edge-state 
channels through a single saddle point in the transition region. Fluctuations of a 
single potential saddle point are related to the Coulomb interaction between localised 
states in the bulk of the disordered 2DES. These localised states are 
treated using the Efros-Shklovskii Coulomb glass model. Models at finite temperature $T$ 
with localised states arranged on regular lattice points or random sites are then 
studied numerically by Monte Carlo methods and exact enumeration. Results for $G$ versus $V_{g}$ at 
different disorder configurations and temperatures are obtained. 
After that, comparisons of our numerical results and experiments are discussed. We conclude with a
summary.

\section{Theoretical model}

Before introducing the model, we note that there are different electron states in a disordered 
two dimensional electron system (2DES) in a high magnetic field. Edge state exists on an 
equi-energy line along the boundary of the sample connecting two contacts. The width of each 
edge-state channel is about the order of magnetic length, $l_{B} \propto 1/{\sqrt{B}}$. Therefore, in 
a high magnetic field the backscattering between edge states at opposite boundaries can be neglected.  
Since each of the edge-state channels is now one-dimensional, we can apply the two-terminal 
Landauer-B{\" u}ttiker formula to calculate the conductance. The current 
injected into each of the edge states is proportional to the 
difference of chemical potentials between two contacts and also each edge-channel pair. In the context 
of the measurements on {\it Si} MOSFETs \cite{cobdenkogan,cobden-prod,cobden-unpublished}, the bias 
voltage between two contacts is fixed, but the gate voltage of the metallic gate on top of the 
2DES, therefore the Fermi energy, is varied. As the number of 
electrons is increased such that the Fermi level is between Landau levels $E_{n}<E_{F}<E_{n+1}$, there 
are $n$ edge channels at each boundary. In a high magnetic field with no inter-edge backscattering, 
the conductance of the system is thus quantised and given by $G=ne^{2}/h$. 

\begin{figure}
\epsfxsize=2.0in
\centerline{\epsffile{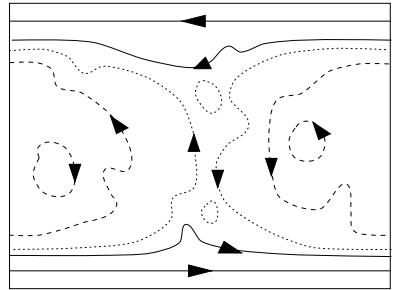}}
\vspace{0.05in}
\caption{The formation of an edge-state channel pair as the gate voltage is increased 
in the transition regime, here between $G=e^{2}/h$ and $2e^{2}/h$. The lines with arrows 
represent the directed trajectories of electrons or holes in the magnetic fields.}
\vspace{0.05in}  
\label{fig:percoln}
\end{figure}

In the transition region between two quantised plateaus, it is clear that the Fermi level is within
a disorder-broadened Landau level. For clarity, let us consider what happens when the conductance 
is between the first and the second plateaus (see Fig.\ \ref{fig:percoln}). 
At zero temperature, there are localised 
states in the bulk associated with closed local equipotential lines \cite{huckstein}, shown as dashed 
and dotted closed lines in Fig.\ \ref{fig:percoln}. Apart from them, 
we can imagine that, in the bulk of the system, there are also directed extended states 
at the boundaries of the regions which are occupied by electrons and connected to one or the other 
contact. At some value of the chemical potential, these extended states undergo large 
excursions into the bulk. As $\mu$ is increased further, 
the directed extended states approach each other more closely and tunneling between them occurs 
(shown as the dotted line in Fig.\ \ref{fig:percoln}). Eventually, a second pair of edge states is 
formed in this way at the inner side near the system boundaries. 

Let us consider the simplest tunneling process in which the two growing regions of 
electrons form a single potential saddle point assumed to be located at about the centre 
of the 2DES. The transmission probability $T_{ij}=|t_{ij}|^{2}$ for the incoming 
channel $i$ and outgoing channel $j$ is already known, for 
the saddle-point potential of the form 
$V(x,y)=V_{0}-(1/2)m\omega_{x}^{2}x^{2}+(1/2)m\omega_{y}^{2}y^{2}$ with 
$\omega_{x}$ and $\omega_{y}$ characterising the shape of the saddle-point potential \cite{fertig}. 
It is given by   
\begin{equation}
T_{ij}=\delta_{ij}\frac{1}{1+{\rm exp}(-\pi\epsilon_{n})} \ ,
\label{eqn:fertig}
\end{equation}
where $\epsilon_{n}=[E-E_{2}(n+1/2)-V_{s-p}]/E_{1}$ with $E$ denoting the total energy 
of the electron and $V_{s-p}=V_{0}$ being the bare potential strength. The two-terminal 
conductance is then given by $G=(e^{2}/h)\sum_{i}T_{ii}$. 

In Eq.\ (\ref{eqn:fertig}), $E_{1}$ and $E_{2}$ are in general complicated functions of $\omega_{c}$ 
(cyclotron frequency), $\omega_{x}$, and $\omega_{y}$. At high magnetic fields, when $\omega_{c}$ 
is much bigger than $\omega_{x}$ and $\omega_{y}$, $\epsilon_{n}$ in Eq.\ (\ref{eqn:fertig}) is 
simply a dimensionless measure 
of the energy of the electron's guiding-centre motion relative to $V_{0}$. For the edge 
state of the percolating region with a guiding centre energy $E_{G}=E_{F}-\hbar\omega_{c}(n+1/2)$, 
the transmission probability is one if $E_{G} \gg V_{0}$. In the other limit that $E_{G} \ll V_{0}$, 
the edge state is completely reflected. The two-terminal conductance as a function of Fermi energy 
at various magnetic-field strengths have been obtained numerically by B{\" u}ttiker \cite{buttiker}. 
Interestingly, although these results are for a 2DES with a point-contact 
constriction or split-gates on top, the same structure is observed in 
a `macroscopic' {\it Si} MOSFET \cite{cobden-unpublished}. From this, 
it seems that considering only one potential saddle-point in the mesoscopic {\it Si} MOSFET 
where measurements were carried out may be reasonable. 

Within the framework of the edge-state picture described above, 
fluctuations of the transmission of percolating states lead to the conductance fluctuations. 
Here the transmission of electron currents depends solely on the potential energy at the saddle point. 
In general, this potential energy should have a contribution from the interactions. 
It is our purpose to relate the interactions to the conductance fluctuations. 
Electrons in our model are separated into categories: those in edge states and those in 
localised states. In principle, a full treatment of two interactions between electrons 
is very complicated \cite{patrick}. For simplicity, 
we neglect interactions amongst electrons in edge states, and just consider interactions 
between different electrons in localised states and their effect on the saddle-point potential. Equally, 
for simplicity, we neglect tunneling processes that mix the states we term localised with each other or 
with those we term edge states. This is correct in a finite sample in the semiclassical limit, when 
for a given value of the Fermi energy tunneling is important only at one saddle-point of
the potential. More generally, one expects states at the plateau transition to be delocalised by 
the tunneling processes we omit, at least within a single-particle description. Note that, although 
interactions between extended electrons are left completely untreated in the model, it turns out, as 
we shall present later, that this way of including the Coulomb interaction does indeed produce dramatic 
fluctuations of the saddle-point potential as the chemical potential of the system is varied. 

The model Hamiltonian describing a system of localised states interacting 
with the Coulomb interaction is essentially the one which was studied first by Efros and 
Shklovskii \cite{coulombgap}. The same Hamiltonian produces a gap in the single-electron 
density of states at the Fermi level due to the Coulomb interaction. 
In the `Coulomb-gap' system, electrons are strongly localised on a discrete set of sites 
due to the impurities. These quantum particles can thus be thought of as being in the 
regime where they behave classically. The Hamiltonian for localised states on
$N$ sites at positions denoted by $i$ is
\begin{eqnarray}
H &=&\sum_{i=1}^{N}{\epsilon_{i}}(n_{i}-\frac{1}{2})+\frac{1}{2}\sum_{i\neq j}^{N}
\frac{e^{2}}{4\pi\varepsilon \varepsilon_{0}}\frac{(n_{i}-\frac{1}{2})(n_{j}-\frac{1}{2})}{r_{ij}} 
\nonumber\\ 
& &-\mu\sum_{i=1}^{N}(n_{i}-\frac{1}{2}),
\label{eqn:coulombglass}
\end{eqnarray}
where $\epsilon_{i}$ is the random site-energy, $n_{i}$ is the occupation number 
which can be either 0 or 1, and $r_{ij}$ is the distance between site $i$ and $j$. 
The subtraction of $1/2$ from $n_{i}$ represents neutralising background charge. 
$\varepsilon$ and $\varepsilon_{0}$ are dielectric constants for the Silicon and 
the vacuum, respectively. The last term in the Hamiltonian is crucial to model the 
experimental variation in the chemical potential in the system as the gate voltage 
is varied. More explicitly, as the voltage of the gate electrode on top of the 2DES 
is varied, there will be electrons brought in or pulled out of the 2DES in the 
inversion layer in the Silicon. The occupancy in 
each of the localised states will then fluctuate as a function of $\mu$, 
due to the Coulomb interaction. Accordingly, the electrostatic potentials between the $N$ localised states and the
 saddle-point potential will also fluctuate.
Hence we have a fluctuating total saddle-point potential 
\begin{equation}
V_{s-p}=\sum_{i}^{N} \frac{e^{2}}{4\pi\varepsilon \varepsilon_{0}}\frac{(n_{i}-
\frac{1}{2})}{\sqrt{r_{i,s-p}^{2}+h^{2}}}-V_{0}
\label{eqn:s-p}
\end{equation}
with $r_{i,s-p}$ representing the distance between site $i$ and the saddle point, 
where $h$ is the perpendicular distance between the inversion layer containing 
localised states and the centre of the saddle point. Again, the occupation at site 
$i$ is given by $(n_{i}-1/2)$ to take into 
account background charges. Note that at finite temperature $T$, in Eq.\ (\ref{eqn:s-p}) 
we need the thermal average of the occupation number, $\langle n_{i} \rangle$, 
instead of $n_{i}$ at each site $i$. As a result, the transmission probability 
given by Eq.\ (\ref{eqn:fertig}) fluctuates as a function of the gate voltage.

\section{Numerical results and comparisons with experiments}

In this section, we present the results obtained numerically at finite temperature.
The 2DES is chosen to have a rectangular shape with length 
$L$ and width $W$ in units of lattice spacing. The localised states are chosen 
to be fixed at sites $i$ arranged on a regular lattice with positions given by 
$(x,y)$. At fixed temperature $T$, we compute $\langle n(x,y) \rangle$, at each site $i$ with 
coordinates ($x,y$) for each chemical potential $\mu$. Note that, here and in all 
the following computation, we have set 
the electric charge, $4\pi\varepsilon \varepsilon_{0}$, and the Boltzmann 
constant $k_{B}$ to be one. For each configuration of impurities, 
a random energy is attached to each localised state. 
The random site-energy $\epsilon_{i}$ at site $i$ is some value in the 
interval $[-{\cal W}/2,{\cal W}/2]$. After ${\langle n(x,y) \rangle}$ is obtained, 
Eq.\ (\ref{eqn:s-p}) determines the value 
of $V_{s-p}$ at this $\mu$ and temperature $T$. The corresponding transmission probability 
is then given by Eq.\ (\ref{eqn:fertig}), rewritten here as 
$T_{ii}=1/{1+{\rm exp}[-(V_{s-p}+c)/E_{0}]}$, 
where $c$ is a constant representing the energy of the electron's guiding 
centre (or the equipotential line) \cite{fertig}; $c$, however, is always set to be equal 
to $V_{0}$ in our numerical study. $E_{0}$ here is an energy scale, chosen 
arbitrarily, which in principle can depend on the magnetic field and the 
characteristics of the saddle-point. The conductances at different chemical 
potentials are  calculated in this way by employing the same set of random 
energies at $N$ sites. For different sets of random site-energies which 
essentially represent different disorder realisations, sample-to-sample 
fluctuations can then be compared. Numerical studies have been carried out using different 
thermal-averaging methods: the Metropolis Monte Carlo algorithm and exact enumeration.

Due to the probabilistic nature of the Metropolis Monte Carlo algorithm, it is important 
to ensure that the system has reached equilibrium, and that the Monte 
Carlo average indeed gives the thermal average. Due to the long-range interaction 
term, we expect that a large system will need an extremely long time to reach equilibrium. 
At higher temperature, stronger disorder ({\it i.e.} larger ${\cal W}$) and larger $|\mu|$, 
this difficulty may be avoided since the contribution of the Coulomb interaction is then small. It 
turns out that at low temperature fluctuations due to lack of 
equilibration are rather large up to the biggest(6$\times$6) system we have reached. In order 
to study the fluctuations due to interactions in more detail, we turn to calculate the 
thermal average using exact enumeration. In doing so, what occurs in lower temperature can 
be investigated more explicitly. 

Exact enumeration means that the thermal averages are achieved by finding explicitly 
all possible state configurations and calculating their Boltzmann factors. 
From the comparison of results using two different methods, 
it is clear that those small-amplitude fluctuations obtained in using the Monte Carlo algorithm
are due to the non-equilibration, instead of the interaction. 

\begin{figure}
\epsfxsize=3.0in
\centerline{\epsffile{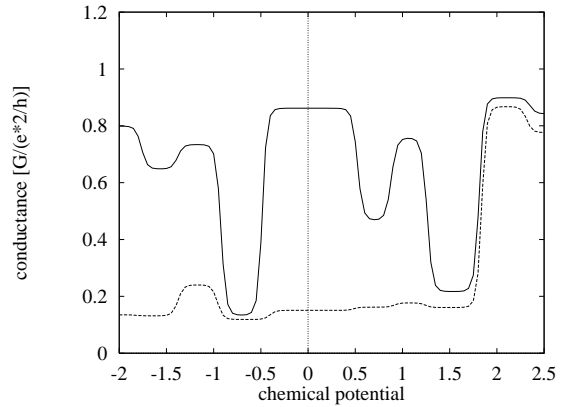}}
\vspace{0.05in}
\caption{Conductance versus chemical potential on a 5$\times$4 lattice at
$T$=0.03, ${\cal W}$=0.2, $E_{0}$=5, and $h$=0.3. The saddle point is chosen to be at 
($x,y$)=(3,2). Two lines are for different realisations of impurity configuration.}
\vspace{0.05in}  
\label{fig:g-mu}
\end{figure}

With reasonable computing time, we obtain results for systems of sizes up to 5$\times$4. 
Here, only the results for 5$\times$4 systems are presented.
Fig.\ \ref{fig:g-mu} clearly exhibits significant conductance fluctuations ranging 
from 0 to $e^{2}/h$ as the chemical potential is varied. The 
amplitudes of the fluctuations, however, depend on the value of the energy 
scale $E_{0}$. We can choose different sequences 
of random numbers in the program to change the random site-energy at each site 
on the lattice. For the same disorder strength ({\it i.e.} ${\cal W}$), the 
sample-to-sample fluctuations due to different realisations of impurities can 
then be studied. In the two cases shown in Fig.\ \ref{fig:g-mu}, there are large 
amplitude fluctuations and the details are sample dependent.

\begin{figure}
\epsfxsize=3.0in
\centerline{\epsffile{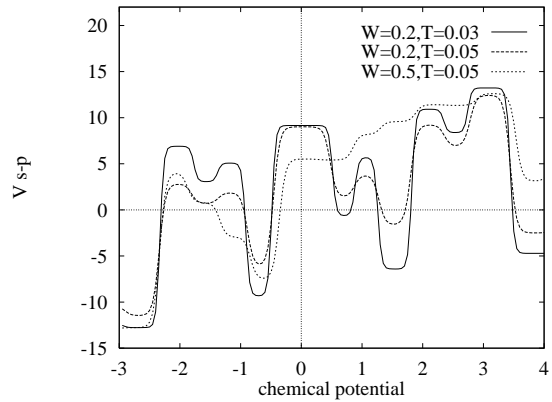}}
\vspace{0.05in}
\caption{ Fluctuations of the saddle point potential, $V_{s-p}$, at ($x,y$)=(3,2) versus 
chemical potential (or gate voltage) on 5$\times$4 systems. $h=$0.3 for all results here.}
\vspace{0.05in}  
\label{fig:5-4}
\end{figure}

To avoid the confusions due to the arbitrary $E_{0}$, only plots of $V_{s-p}$ versus 
$\mu$, instead of $G$ versus $\mu$, are shown in Fig.\ \ref{fig:5-4} where $h$ is the
only free parameter to obtain $V_{s-p}$. We then compare results at different 
disorder strengths. Fluctuations due to the Coulomb interaction should be suppressed as 
the strength of the disorder is increased. This is more or less consistent 
with what is demonstrated in Fig.\ \ref{fig:5-4}. Finite temperature is 
another source in our model which can smear out the amplitudes of the 
fluctuations. This is because, with a fixed chemical potential, different thermal energies 
give different averaged occupation numbers on the same lattice site. At higher temperature,
each electron gains more thermal energy on average and thus the effect of Coulomb interaction 
is again suppressed. The temperature dependence of the solid and dashed lines 
shown in Fig.\ \ref{fig:5-4} with the same disorder strength clearly shows this behaviour. 

Finally, we discuss the results on a system which 
contains randomly distributed localised states. This seems to be a more realistic 
representation of real devices. We use the exact enumeration for the thermal averaging. The 
positions of the localised states are chosen independently 
with a uniform distribution over the system. For the same number 
of localised states as that in the lattice system, we conclude that 
finite-size effects are stronger than on the lattice. Although
not shown here, results for different disorder strengths show no significant fluctuations. 
Comparing with the results for the lattice system, fluctuations 
are much smaller. In order to obtain prominent interaction-driven 
fluctuations off lattice, it seems possible that we need to study very large random-site systems. 

We now turn to describe the comparison between the numerical 
results from our model on a lattice and the experiments. First, let 
us focus on the conductance as a function of the gate voltage at some 
fixed magnetic field. The conductance fluctuations produced from our 
model are sample dependent, and varying the saddle-point potential 
with chemical potential can produce conductance fluctuations between 
$G$=0 and $e^{2}/h$. Both of these features are consistent with what has 
been observed in experiments \cite{cobdenkogan,cobden-prod,cobden-unpublished}. However, 
the fluctuation patterns are rather different for our model and the 
experiments. It is observed in experiments that the conductance fluctuates 
in sharp peaks and dips in the transition region between two plateaus 
\cite{cobdenkogan,cobden-prod,cobden-unpublished}. 
By contrast, at least up to the system sizes for which we have done the computation, it is 
observed that, from Fig.\ \ref{fig:g-mu} and \ref{fig:5-4}, sharp spikes of the fluctuations 
are obviously absent in our results. Instead, steps with rounded or flat 
tops or bottoms occur as we vary $\mu$. This feature can be understood in 
terms of the total electrostatic potential energy changing slowly or even 
remaining constant for some finite interval of the chemical potential. It is 
also true that the systems we have 
studied show a relatively small number of conductance fluctuations as $\mu$ 
varies, compared with experiments. However, we should expect more fluctuations 
in a bigger system. Our model seems to show this behaviour as the system size is
increased. It is therefore not appropriate to analyse the distribution 
of conductance \cite{cobdenkogan} from our results which only contain small 
numbers of independent conductance values.    

Another important comparison is the temperature dependence of the conductance 
fluctuations. Finite temperature reduces the amplitudes of the fluctuations because 
the fluctuation of the occupation number in each localised state is smoothed out 
by the finite thermal energy. In our model, it can be observed that the width of 
each fluctuation `steps' does not vary with temperature 
(see Fig.\ \ref{fig:5-4}). In experiments, data \cite{cobden-unpublished} demonstrate 
indeed that the amplitudes of peaks and dips are enhanced as the temperature is lowered. 
No obvious shrinking of their widths can be seen there.     

We now come to the behaviour of fluctuation peaks and dips at different magnetic 
fields. Although there is no explicit magnetic-field dependence in our model, we 
argue here, by noting that straight lines on 
the $V_{g}$-$B$ plane \cite{cobden-unpublished} are obtained provided the system preserves 
the filling factor along each line, the occurrence of straight lines is consistent with 
the spirit of our model.  

As mentioned in the introduction section, the slopes of two sets of 
lines are parallel to those associated with the centres of the neighbouring
plateaus. More explicitly, it is actually 
observed that each of these straight lines follows the equation 
$V_{g}=CB+D$ with $C$ and $D$ being constants. This behaviour 
can be understood in terms of the physics of the {\it Si} MOSFET \cite{andofowlerstern},
as has been discussed in Ref. 23. 
With the filling factor given by the relation 
$\nu=(V_{g}-V)\varepsilon\varepsilon_{0}h/de^{2}B$ 
as a perpendicular magnetic field $B$ is applied, we have 
\begin{equation}
V_{g}=\frac{e^{2}d}{\varepsilon\varepsilon_{0}h}\nu B+V.
\label{eqn:slope}
\end{equation} 
Here $V$ is some constant and represents a threshold voltage, and $d$ is a distance of the 
order of the thickness of the $SiO_{2}$ layer. To describe the $i$th straight line on the 
$V_{g}$-$B$ plot, 
another constant $V_{i}$, for example, is needed. For different parallel lines on the 
$V_{g}$-$B$ plane, we have different $V_{i}$'s. Along each line, the filling factor and $V_{i}$ 
are constants. This means that, as $V_{g}$ and $B$ are both varied along the line, the 
fluctuation peaks and dips evolve in such a way that 
the filling factor of the system is unchanged for a given fluctuation. 
More explicitly, the value of $\nu$ is observed to be either $i$ or $i+1$ 
for the $(i+1)$th transition region, depending on which plateau region the 
lines belong to. 

In our model, conductance fluctuations are associated with 
the occupancy of localised states in the 2DES. For the occupancy to fluctuate, 
the state must have energy near the chemical potential. Hence, along each 
straight line, which connects peaks or dips for different $(V_{g},B)$, the 
localised states are at the chemical potential. The fact that $\nu$ along the 
straight line is the same as $i$, for example, associated with the $i$th plateau 
centre means that the localised states in both cases must have the same total 
(kinetic plus electrostatic) energy as the states in the $i$th Landau level. 
This is indeed possible if the localised states belong to the Landau levels of the 
the 2DES. Following this argument, as the chemical potential crosses the centre of a 
disordered-broadened Landau level in the transition region, localised states at two tails of 
Landau level then give the two slopes of straight lines corresponding to two different 
energies of the states at two plateau centres.

\section{Conclusions}

In this paper, we have constructed a simple semiclassical model which produces conductance 
fluctuations in strong magnetic fields due to the Coulomb interaction between 
electrons. The inclusion of the long-range Coulomb interaction causes the occupation 
number in each localised state in the bulk to fluctuate as the chemical potential is 
varied. By taking into account the influence of the electrostatic potential between 
localised states and the saddle point, the energy of the saddle point in the potential 
seen by mobile electrons then also varies with the chemical potential. Through this 
saddle point, the conductance due to the transmission of edge states from one 
contact to the other thus fluctuates with the chemical potential. We study the model 
at finite temperature by numerical simulation using the Monte Carlo methods and exact 
enumeration. At low temperature, the Monte Carlo results suffer slow equilibration. 
Strong fluctuations due to the non-equilibration of the system in this case make it 
difficult to extract interaction-driven fluctuations. By contrast, results obtained 
using exact enumeration clearly exhibit significant fluctuations as a function of the 
chemical potential. 

In comparing these results with the experiments, our model shows qualitatively 
consistent behaviours with the experiments as the gate voltage and the magnetic 
field are both varied. There are, however, some different features existed between 
our results and experiments. In particular, although our simulations indeed exhibit 
fluctuations depending on realisations of disorder they give fluctuations which are 
like steps instead of the sharp peaks and dips observed in experiments. 
These discrepancies could arise because we have neglected interactions between bulk 
extended electrons and many specific details in the {\it Si} MOSFET. 

\section*{acknowledgements}
The author is greatly indebted to Dr. John Chalker for numerous 
discussions. Special thanks to Dr. David Cobden for discussions and  
providing his experimental results before publishing, Dr. Derek Lee for the
help of the programming, and Dr. Chi-Te Liang for the consultation of the 
general experimental details. This work was supported in part by the ORS Award from the 
CVCP in United Kingdom.






\begin{references}
\bibitem{todai} Present address: Department of Applied Physics, 
University of Tokyo, 7-3-1, Hongo, Bunkyo-ku, Tokyo 113, Japan. 

\bibitem{ucf-exp} S.B. Kaplan and A. Hartstein, Phys. Rev. Lett. {\bf 56}, 2403 (1986); 
W.J. Skocpol, P.M. Mankiewich, R.E. Howard, L.D. Jackel, D.M. Tennant, and A.D. Stone, 
Phys. Rev. Lett. {\bf 56}, 2865 (1986).

\bibitem{washburn} S. Washburn, IBM J. Res. Develop. {\bf 32}, 335 (1988); for a list of 
related references, see S. Das Sarma, T. Kawamura, and S. Washburn, Am. J. Phys. {\bf 63}, 
683 (1995).

\bibitem{ucf} P.A. Lee, A.D. Stone, and H. Fukuyama, Phys. Rev. B {\bf 35}, 1039 (1987); 
C.W.J. Beenakker and H. van Houten, in {\it Solid State Physics, vol. 44}, eds. H. Ehrenreich and
D. Turnbull ( Academic Press, New York, 1991).


\bibitem{halperin} B.I. Halperin, Phys. Rev. B {\bf 25}, 2185 (1982).

\bibitem{simmons} J.A. Simmons, S.W. Hwang, D.C. Tsui, H.P. Wei, L.W. Engel, and
M. Shayegan, Phys. Rev. B {\bf 44}, 12933 (1991).

\bibitem{maingeim} P.C. Main, A.K. Geim, H.A. Carmona, C.V. Brown, T.J. Foster,
R. Taboryski, and P.E. Lindelof, Phys. Rev. B {\bf 50}, 4450 (1994).

\bibitem{geim} A.K. Geim, P.C. Main, P.H. Beton, L.Eaves, S.P. Beaumont, and C.D.W. Wilkinson, 
Phys. Rev. Lett. {\bf 69}, 1248 (1992); C.V. Brown, A.K. Geim, T.J. Foster, C.J.G.M. Langerak, 
and P.C. Main, Phys. Rev. B {\bf 47}, 10935 (1993).

\bibitem{SdH} A. Morgan, D.H. Cobden, M. Pepper, G. Jin, Y.S. Tang, and
C.D.W. Wilkinson, Phys. Rev. B {\bf 50}, 12187 (1994).

\bibitem{jainkivelson} J.K. Jain and S.A. Kivelson, Phys. Rev. Lett. {\bf 60}, 1542 (1988).


\bibitem{kinaret} J.M. Kinaret and P.A. Lee, Phys. Rev. B {\bf 43}, 3847 (1991).

\bibitem{dima} D.E. Khmel'nitskii and M. Yosefin, Physica A {\bf 200}, 525 (1993); 
{\it ibid}, Surf. Sci. {\bf 305}, 507 (1994); D.L. Maslov and D. Loss, Phys. Rev. Lett. 
{\bf 71}, 4222 (1993); S. Xiong, N. Read, and A.D. Stone, Phys. Rev. B {\bf 56}, 3982 (1997).

\bibitem{huckstein} B. Huckestein, Rev. Mod. Phys. {\bf 67}, 357 (1995).

\bibitem{cobdenkogan} D.H. Cobden and E. Kogan, Phys. Rev. B {\bf 54}, R17316 (1996).

\bibitem{andofowlerstern} For general aspects of MOSFETs, see, for example, T. Ando, A.B. Fowler, 
and F. Stern, Rev. Mod. Phys. {\bf 54}, 437 (1982).


\bibitem{ando} T. Ando, Phys. Rev. B {\bf 49}, 4679 (1994).

\bibitem{wangjovanovic} Z. Wang, B. Jovanovi{\'c}, and D.-H. Lee, Phys. Rev. Lett. {\bf 77}, 
4426 (1996).

\bibitem{chofisher} S. Cho and M.P.A. Fisher, Phys. Rev. B {\bf 55}, 1637 (1996).

\bibitem{cf-network} B. Jovanovi{\'c} and Z. Wang, Phys. Rev. Lett. {\bf 81}, 2767 (1998).

\bibitem{avishai}Y. Avishai, Y. Band, and D. Brown, Report No. cond-mat/9901328. 

\bibitem{galstyan} A.G. Galstyan and M.E. Raikh, Phys. Rev. B {\bf 56}, 1422 (1997).

\bibitem{cobden-prod} D.H. Cobden, C.H.W. Barnes, C.J.B. Ford, J.T. Nicholls, and M. Pepper, in 
{\it Proceedings of the 24th ICPS}, ed. D. Gershoni (World Scientific, Singapore, 1999).

\bibitem{cobden-unpublished} D.H. Cobden, C.H.W. Barnes, and C.J.B. Ford, Phys. Rev. Lett. {\bf 82}, 
4695 (1999); D.H. Cobden, private communications.


\bibitem{fertig} H.A. Fertig and B.I. Halperin, Phys. Rev. B {\bf 36}, 7967 (1987); 
M. B{\" u}ttiker, in {\it Semiconductors And Semimetals, Vol. 35}, ed. M.Reed (Academic Press, 
San Diego, 1992).

\bibitem{buttiker} M. B{\"u}ttiker, Phys. Rev. B {\bf 41}, 7906 (1990).

\bibitem{patrick} See discussions, for example, in P.A. Lee and T.V. Ramakrishnan, 
Rev. Mod. Phys. {\bf 57}, 287 (1985).

\bibitem{coulombgap} A.L. Efros and B.I. Shklovskii, J. Phys. C: Solid State Phys. {\bf 8}, 
L49 (1975); A.L. Efros and B.I. Shklovskii, in {\it Electron-Electron Interaction in Disordered Systems}, 
eds. A.L. Efros and M. Pollak (North-Holland, Amsterdam, 1985).



\end{references}
\end{document}